\documentclass[lettersize,journal]{IEEEtran}
\usepackage{amsmath,amsfonts}
\usepackage{algorithmic}
\usepackage{array}
\usepackage[caption=false,font=normalsize,labelfont=sf,textfont=sf]{subfig}
\usepackage{textcomp}
\usepackage{stfloats}
\usepackage{url}
\usepackage{verbatim}
\usepackage{booktabs}
\usepackage{arydshln}
\usepackage{xcolor}
\usepackage{listings}
\usepackage{arydshln}
\usepackage{multirow}
\usepackage{musixtex}
\usepackage{hyperref}
\usepackage{textcomp}
\usepackage{threeparttable}
\usepackage{stmaryrd}

\usepackage{graphicx}
\def\BibTeX{{\rm B\kern-.05em{\sc i\kern-.025em b}\kern-.08em
    T\kern-.1667em\lower.7ex\hbox{E}\kern-.125emX}}
\usepackage{balance}

\newcommand{\review}[1]{#1}

\begin{document}
\title{Can Audio Reveal Music Performance Difficulty? Insights from the Piano Syllabus Dataset}

\author{
\IEEEauthorblockN{Pedro Ramoneda\IEEEauthorrefmark{1}, Minhee Lee\IEEEauthorrefmark{2}, Dasaem Jeong\IEEEauthorrefmark{2}, Jose J. Valero-Mas\IEEEauthorrefmark{3}, Xavier Serra\IEEEauthorrefmark{1}}

\IEEEauthorblockA{\IEEEauthorrefmark{1}\textit{Music Technology Group, Universitat Pompeu Fabra, Spain}}

\IEEEauthorblockA{\IEEEauthorrefmark{2}\textit{Music \& Art Learning Lab, Sogang University, Korea}}

\IEEEauthorblockA{\IEEEauthorrefmark{3}\textit{Pattern Recognition and Artificial Intelligence Group, University of Alicante, Spain}
\vspace{-0.6cm}}

\thanks{\IEEEauthorrefmark{1}Corresponding author: pedro.ramoneda@upf.edu}
}

\maketitle

\begin{abstract}
Automatically estimating the performance difficulty of a music piece represents a key process in music education to create tailored curricula according to the individual needs of the students. Given its relevance, the Music Information Retrieval (MIR) field \review{comprises} some proof-of-concept works addressing this task that mainly focus on high-level music abstractions such as machine-readable scores or music sheet images. \review{In this regard, the potential of directly analyzing audio recordings has generally been neglected}. This work addresses this gap in the field with \review{two contributions}: \review{(i) PSyllabus, the first audio-based difficulty estimation dataset---collected from \emph{Piano Syllabus} community---}featuring 7,901 piano pieces across 11 difficulty levels from 1,233 composers as well as two additional benchmark datasets particularly compiled for evaluation purposes; and (ii) a recognition framework capable of managing different input representations---both in unimodal and multimodal manners---derived from audio to perform the difficulty estimation task. The comprehensive experimentation comprising different pre-training schemes, input modalities, and multi-task scenarios proves the validity of the \review{hypothesis} and establishes PSyllabus as a reference dataset for audio-based difficulty estimation in the MIR field. The dataset, developed code, and trained models are publicly shared to promote further research in the field.
\end{abstract}


\section{Introduction}
\review{In the education field, estimating the performance difficulty of a music piece enables} the \review{personalized design} of the learning curriculum of a student~\cite{curriculumdesign,mellizo2020music}. \review{However, the process of manual difficulty estimation is inherently laborious and time-consuming, and while subjective elements may arise in fine-grained details, professionals generally agree on coarse scope.} \review{This demanding process also contributes to the rigidity of curricula in many music schools and conservatoires~\cite{cutietta2007content,foley2024modernizing}, as the exploration and \review{compilation} of new repertoire from vast collections is often difficult and resource-intensive.}

\begin{figure}[ht!]
  \centering
  \includegraphics[trim=1.4cm 3cm 4.3cm 2cm, clip, width=.99\linewidth]{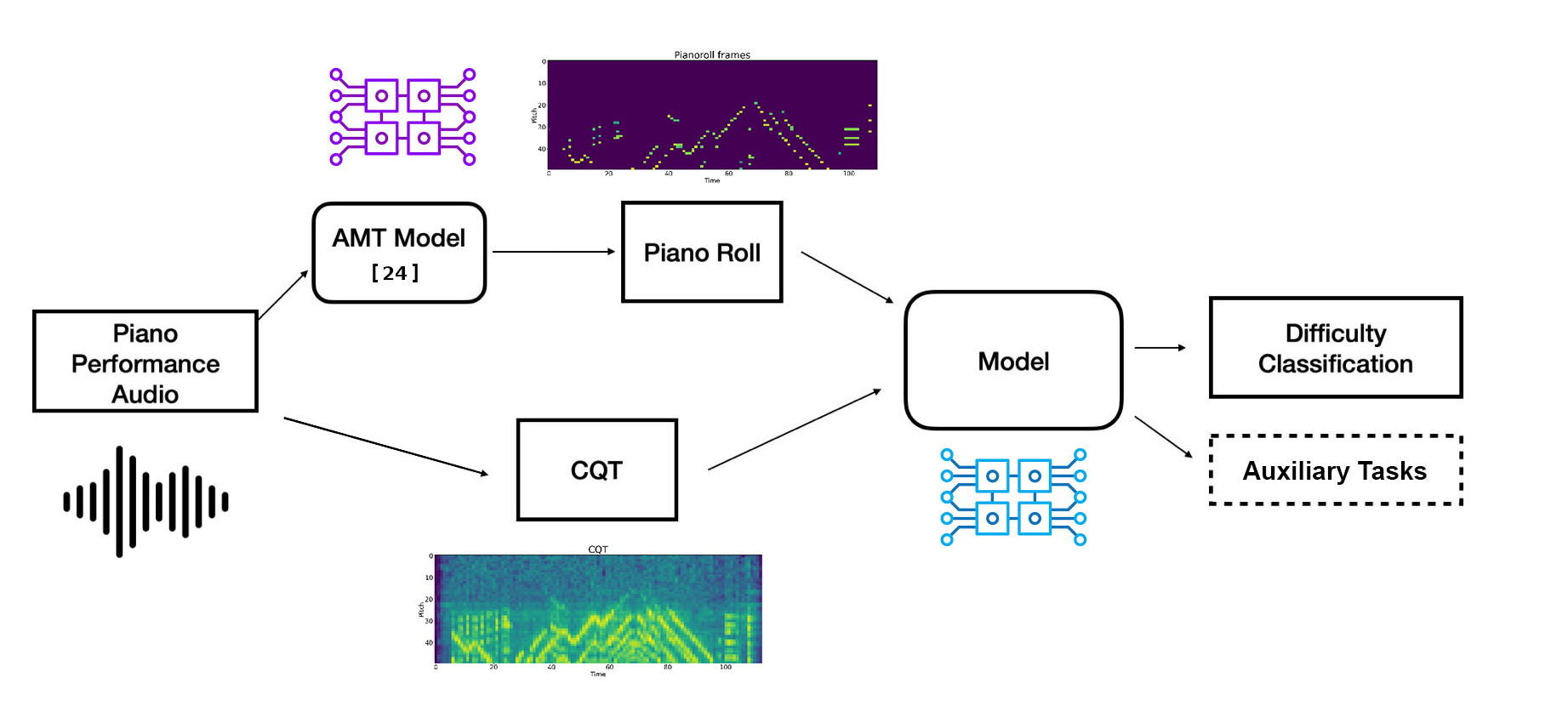}
  \vspace{-0.5cm}
  \caption{We introduce a recognition framework that utilizes both piano-roll and Constant-Q Transform audio-derived representations to estimate the difficulty of a given piece. \review{The model is trained on the novel PSyllabus dataset using distinct configurations (unimodal and multimodal)} and explores multiple training strategies, including multi-tasking with auxiliary tasks, offering valuable insights into the task.
  }
  \label{fig:teaser}
  \vspace{-0.7cm}
\end{figure}

In response to this, researchers in the Music Information Retrieval (MIR) field have devoted considerable efforts for over a decade to automatize and systematize this difficulty estimation process~\cite{chiu2012study,sebastien2012score,nakamura2018statistical}. While still an open \review{problem} in MIR~\cite{deconto2023automatic, ramoneda2022}, these advances have not only found application in the automatic organization of music repertoires, but they have also enabled the discovery and promotion of music pieces ignored throughout history~\cite{ramoneda2023predicting}. Additionally, these advancements have led to the development of automatic rearrangement systems capable of adapting the difficulty of a given piece to a different target one~\cite{gover2022music,suzuki2023piano,nakamura2014merged}. Finally, it must be highlighted that the interest in this field transcends academia and reaches the industry sector, since relevant music-oriented companies such as Muse Group\footnote{\url{https://musescore.com/}, \url{https://www.ultimate-guitar.com/}} or Yousician~\cite{kaipainen2017system} have also proposed commercial systems for performing this task.

Research on difficulty estimation has predominantly focused on piano music~\cite{sebastien2012score,ramoneda2022,ramoneda2024}, using symbolic machine-readable scores (e.g., MusicXML format)~\cite{nakamura2015automatic,zhang2023symbolic} or sheet images~\cite{ramoneda2023predicting} as sources of information. \review{Most solutions frame this task as a classification problem, but some focus on analysis~\cite{chiu2012study} or use rule-based methods, like Nakamura et al.~\cite{nakamura2014merged}, which relate piano fingering patterns with difficulty levels. Nakamura et al.~\cite{nakamura2015automatic} later extended these ideas for music generation using hidden Markov models. Recent work has also explored new directions, such as Vasquez et al.~\cite{vasquez2023quantifying}, focusing on guitar, and Ramoneda et al.~\cite{ramoneda2024}, applying transfer learning from piano fingering and expressiveness models.}

However, attending to the multimodal nature of music~\cite{schedl2014music}, we may argue that current development of the difficulty estimation field is severely biased towards notated music---either symbolic or \review{score images}---, which remarkably limits its reach~\cite{zhang2024audio}. Based on this premise, this work proposes a paradigm shift towards the analysis of \review{audio} music recordings for several reasons: from a pedagogical perspective, this development should enable students to easily explore a wider range of music---e.g., streaming platforms---, encouraging them to actively discuss and participate in their learning preferences, and increasing their motivation and engagement in the learning process; from an MIR perspective, this paradigm shift should provide additional insights to those obtained in notated music, hence expanding the knowledge related to the field.

Due to the novelty of the formulation, dealing with acoustic recordings poses a novel set challenges to this difficulty estimation task: (i) the lack of annotated datasets depicting acoustic recordings with their difficulty level; (ii) obtaining a suitable representation of the audio signal that allows for its adequate analysis; and (iii) devising an appropriate \review{deep-learning framework} capable \review{of} dealing with different representations---individually as an \textit{unimodal} model as well as concurrently in a \textit{multimodal} fashion---of the \review{audio} music recording, as illustrated in Fig.~\ref{fig:teaser}. We now briefly describe these issues as they constitute the core aspects addressed in the manuscript.

To \review{alliviate} the data scarcity issue, \review{we carefully curated and cleaned the data to ensure its quality and usability. The data was sourced from the Piano Syllabus community, which, to the best of our knowledge, has never been used for machine learning purposes.}
\footnote{Available at: \url{https://pianosyllabus.com/}.}
After a semi-automatic cleansing process, the compiled dataset comprises 7901 pieces from than 1233 different composers spanning along 11 difficulty labels. Additionally, we introduce two small benchmark datasets: (i) a first assortment, which extends the one introduced by Ramoneda et al.~\cite{ramoneda2024}, that comprises 42 audio pieces by black women composers from the Hidden Voices Project~\cite{hidden} categorized by difficulty; and (ii) an additional dataset that consists of multiple performances of 55 representative pieces of each difficulty level and music period to assess \review{whether a decision-level fusion of the individual estimates for different performances of a music piece may report more robust performance figures than the stand-alone estimation case.}

The second challenge consists in deriving a suitable representation that allows for direct analysis, akin to the challenges faced with sheet music and symbolic music representations. For that, one may resort to the Automatic Music Transcription field to obtain a symbolic representation of the music content in the acoustic recording~\cite{benetos2018automatic}, being this premise particularly supported by the remarkable advancements that this field has recently experienced in piano transcription~\cite{kwon2024towards,transcription2022kong}. Nevertheless, since the direct use of spectral features still stands as a competitive alternative in areas such as computational musicology~\cite{rege2020review} or automated classification~\cite{alonso2022music}, this work studies and compares these two modalities.

In relation to the learning framework, we resort to the so-called Convolutional-Recurrent Neural Network (CRNN) scheme complemented with attention mechanisms due to its reported competitiveness in the related fields of music classification and tagging~\cite{Choi:ICASSP:2017,Won:SMC:2020}. \review{More specifically}, this architecture comprises an initial convolutional stage to extract local acoustic features that are then summarized using a recurrent model complemented with attention mechanisms to focus on relevant parts of the data sequence for the task at hand~\cite{Srivastava:CICT:2022, Won2021SemisupervisedMT}. As previously discussed, \review{This scheme is designed to be trained and evaluated in both unimodal and multimodal configurations. Each configuration uses specific representations of the acoustic signal, either individually (e.g., CQT or piano-roll) or in combination, but always as separate model instances.} Moreover, similarly to recent proposals from the music classification field~\cite{Wu:ICASSP:2021}, we also assess the applicability of multi-task learning---i.e., the process of training the model to solve multiple tasks altogether for improving its generalization capabilities~\cite{zhang2021survey}---for our task, being the additional processes considered those of composer recognition, music era/period classification, and difficulty ranking estimation.

Considering all the above, our \review{contributions} are:
\begin{itemize}
    \item We adopt a CRNN model with attention mechanisms to estimate the performance difficulty of piano recordings.
    \item We introduce the novel PSyllabus data collection of acoustic piano recordings with their associated difficulty levels suitable for training deep learning models. 
    \item We propose two additional benchmark-oriented music collections: (i) an assortment comprising relatively unknown composers to foster research on underrepresented communities in MIR; and (ii) a multi-performance assortment that accounts for researching how multiple performances influence the difficulty prediction.
    \item We conduct extensive experiments to assess the proposed methodologies, including a zero-shot scenario for testing generalization in \review{out-of-distribution} cases, multi-task learning with auxiliary tasks, and training with multiple difficulty rankings.
    \item To promote the task in the MIR community as well as for reproducible research, the code, models, and the datasets are publicly available for research-oriented purposes.\footnote{Dataset: \url{https://doi.org/10.5281/zenodo.14794592},\\ GitHub: \url{https://github.com/pramoneda/audio-difficulty}.}
\end{itemize}

\lstset{
    basicstyle=\scriptsize\ttfamily, 
    breaklines=true, 
    breakatwhitespace=false, 
    escapeinside={(*@}{@*)}, 
}


The rest of the work is structured as follows: Section~\ref{dataset} introduces the PSyllabus dataset and benchmark collections; Section~\ref{methodology} details the proposed methods; Section~\ref{sec:experimental_setup} describes the experimental setup and neural architecture; Section~\ref{sec:results} presents results on input representation and training; Sections~\ref{sec:case_women} and~\ref{case_multi} cover case studies on gender bias and model evaluations, respectively; and Section~\ref{conclusions} wraps up the work and suggests future directions.

\vspace{-0.2cm}
\section{The Piano Syllabus Dataset}
\label{dataset}

\begin{figure*}[ht!]
\begin{lstlisting}
(*@\textcolor{red}{Request to API: }@*)
    You are an expert in classical piano repertoire.

    Task: Given two piano piece titles, A and B, determine if A is the same piece as B, or if A is a movement or part of B, or the reverse.

    Consider that a piece might have multiple movements. Title B is from a YouTube video. If B is a piece with multiple movements and A is a movement or part of B, say A is a partial of B, or the opposite.

    Say "The same piece" if A is exactly the same as B and "Distinct piece" if they are not related pieces.
    
    Example:
    A: Bach J.S. Partita No. 1 in Bb major BWV 825 - Allemande
    B: J.S. Bach Partita no. 1 in B-flat Major, BWV 825 - Chelsea Wang, piano
    Answer: A is a single movement of B

    ... FIVE MORE EXAMPLES ...

    ------------- Answer the questions below, briefly. Keep it short like the provided examples. ------
    A: Shostakovich D. - Prelude \& Fugue No 23 in F major Op 87
    B: Shostakovich D: Preludes and Fugues for Piano, Op. 87 - Prelude & Fugue No. 23 in F major: Fugue

(*@\textcolor{red}{Answer: }@*)
    B is a partial of A

\end{lstlisting}
\vspace{-0.3cm}
\caption{Prompt engineering template for ChatGPT (version 4) used to validate the consistency of the PSyllabus dataset.}
\label{fig:chatgpt}
\vspace{-0.5cm}
\end{figure*}

\begin{figure}[ht!]
\centering
\includegraphics[trim=0 0.7cm 0 1.4cm 0, clip,width=.99\linewidth]{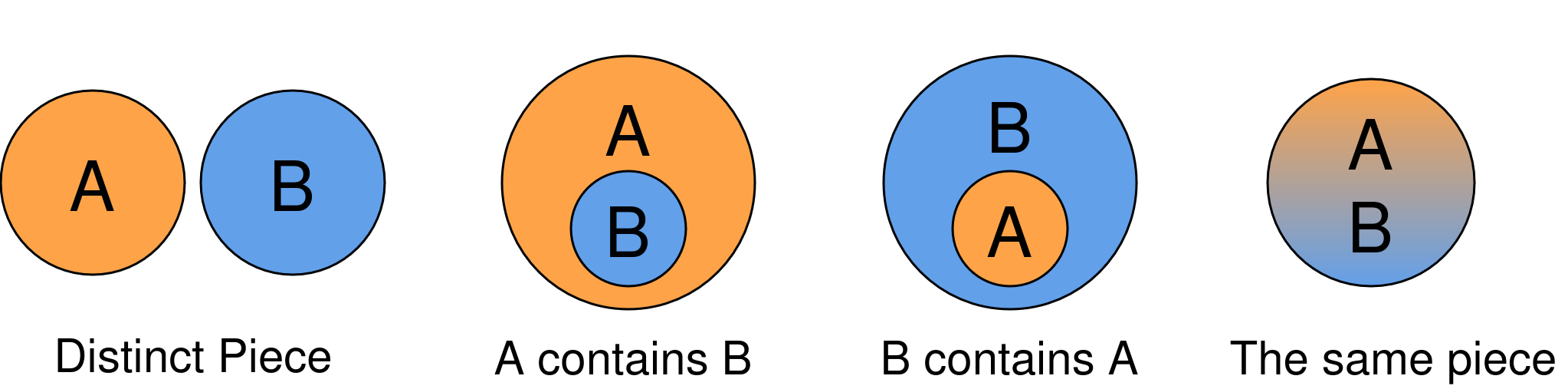}
\caption{Possible scenarios in the metadata of two given music pieces before the prompt engineering filtering stage.}
\label{fig:sets_diagram}
\vspace{-0.7cm}
\end{figure}

Created in 2014 by Ian Wheaton, Piano Syllabus is a web community dedicated to compiling the largest syllabus of piano compositions ranked by performance difficulty. This project has collected over 12\,000 pieces and categorized them into 11 levels based \review{attending to the performing difficulties of their corresponding scores}. Our research aims to leverage this domain knowledge to develop deep learning models for automatically estimating the difficulty level of piano performance recordings. We introduce the PSyllabus dataset of audio recordings and difficulty levels, compiled from the aforementioned web resource. This dataset is distributed—including \review{links to all audio recordings and metadata used for its compilation}—for research-only purposes under a \review{Creative Commons 4.0 license}. \review{Data is available upon requests made via the Zenodo platform}. The license explicitly prohibits \review{any commercial use, redistribution of audio content, or attempts to reverse-engineer precomputed representations}. \review{Transcribed MIDI files and CQT representations are provided to support research reproducibility within these limitations}.

The remainder of the section delves into the methodology to create the dataset (Section~\ref{ss:dataset_creation}), then describes and analyzes its features (Section~\ref{ss:dataset_analysis}), and finally introduces the two additional benchmark datasets (Section~\ref{benchmark_datasets}).

\subsection{Compilation and curation}
\label{ss:dataset_creation}

The first task in creating the dataset was collecting acoustic recordings of the pieces in the Piano Syllabus resource. Out of the initial 12\,000 compositions indexed in this database, only 9\,829 pieces could be retrieved from the Internet, \review{with YouTube as the sole source}. \review{Once retrieved, all pieces were resampled to a resolution of 44.1 kHz.}

After that, this assortment underwent a curation process to address different issues found: (i) low-quality recordings; (ii) \review{out-of-scope pieces (e.g., duets, chamber music, etc)}; and (iii) multiplicity in the difficulty grades (i.e., different difficulty levels assigned to the same piece). This filtering process resulted in 8\,426 pieces out of the 9\,829 initially gathered.

In addition to the previously highlighted points, \review{research on \textit{Piano Syllabus} collection faces a key issue: annotations sometimes assign a single difficulty level to an entire recording, even when the difficulty varies within it. For example, \textit{Shostakovich's 24 Preludes Op.~34} includes pieces rated between levels 7 and 10, but all share a link to the same recording, creating inconsistencies during model training. Conversely, some annotations for entire sonatas are linked to single movement videos, leading to mismatched labels. Both scenarios complicate the model's ability to learn accurate difficulty representations.}

To address this issue, we devised a validation approach based on LLMs. More precisely, we resorted to ChatGPT (version 4) to validate the consistency of the annotations in the music pieces via prompt engineering. \review{The goal in this case was to assess whether, for two given pieces, one was part of the other, whether they were distinct compositions, or whether they were the same piece. As illustrated in Fig.~\ref{fig:sets_diagram}, only the pieces identified as the same were maintained in the collection.}
An example of such process is shown in Fig.~\ref{fig:chatgpt}.

Based on this LLM-based process, 525 pieces were discarded due to containing inconsistent annotations, hence comprising the resulting PSyllabus dataset 7\,901 pieces.

\subsection{Analysis and statistics}
\label{ss:dataset_analysis}

\review{This section analyses the main features of the PSyllabus dataset. Table~\ref{tab:dataset} describes this collection and compares it against the aforementioned datasets from the difficulty estimation literature, in terms of the total number of pieces, difficulty classes, and composers. The \textit{average imbalance ratio} (AIR), obtained as the mean of the individual ratios between each difficulty class and the majority label in the collection~\cite{ramoneda2023predicting}, is also provided to assess the imbalance degree of the datasets.}


\begin{table}[h!]
\centering
\caption{Statistics of existing datasets for piano performance difficulty estimation.}
\resizebox{.95\columnwidth}{!}{%
\begin{threeparttable}
\begin{tabular}{llcccc}
\toprule[1pt]
\multicolumn{2}{l}{\textbf{Dataset}} & \textbf{Pieces} & \textbf{Classes} & \textbf{AIR} & \textbf{Composers}\\ 
\cmidrule(lr){1-6}
\multicolumn{2}{l}{\textit{Symbolic data}}\\
& MK~\cite{ramoneda2022} & 147 & \review{3} & .78 & 1\\
& CIPI~\cite{ramoneda2024} & 652 & 9 & .33 & 29\\
\hdashline
\multicolumn{2}{l}{\review{\textit{Score images}}}\\
& PS~\cite{ramoneda2023predicting} & 2816 & 9 & .24 & 92\\
& FS~\cite{ramoneda2023predicting} & 4193 & 5 & .37 & 747\\
& HV~\cite{ramoneda2023predicting} & 17 & 4 & .35  & 10\\
\hdashline
\multicolumn{2}{l}{\textit{Audio recordings}}\\
& PSyllabus & 7901 & 11 & \review{$\sim 1$} & 1233\\
& HV Audio\textsuperscript{$\flat$} & 57 & 7 & .43 & 23\\
& Multiple Performances\textsuperscript{$\flat$} & 550 & 11 & \review{1} & 35\\
\bottomrule[1pt]
\end{tabular}
\begin{tablenotes}
\item[$\flat$] Benchmark datasets
\end{tablenotes}
\end{threeparttable}
}
\label{tab:dataset}
\end{table}

\begin{figure}[ht!]
\centering
\includegraphics[trim=0pt 0pt 0pt 5pt, clip, width=.95\columnwidth]{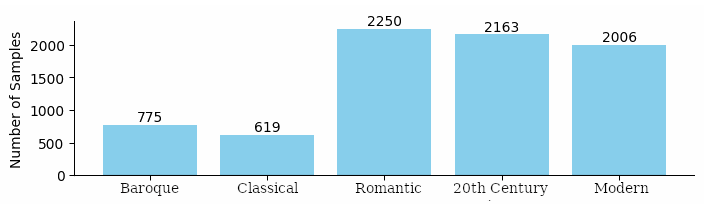}
\caption{Era distribution of PSyllabus dataset.}
\label{fig:era_distro}
\end{figure}

The PSyllabus dataset shows the largest number of pieces among the existing datasets, the highest diversity in terms of composers, as well as one of the collections with the \review{largest amount of difficulty levels}. Moreover, the PSyllabus dataset exhibits an average imbalance ratio $\mbox{AIR}\approx1$, which indicates a balanced distribution across its classes. \review{In comparison, the HV Audio dataset shows the highest imbalance level ($\mbox{AIR}=0.43$) of the presented collections.\footnote{\review{While this labeling bias may be detrimental for a recognition model, its analysis may provide music insights beyond the difficulty level.}} On the other hand, the Multiple Performances dataset achieves perfect balance, as it includes exactly five recordings for each style at every difficulty level. Note that, this latter collection shows no overlap with the PSyllabus dataset.}

Focusing now on the metadata of the pieces, Fig.~\ref{fig:era_distro} illustrates the distribution of musical pieces across different historical eras of the PSyllabus collection.  \review{The classification of pieces by historical era follows the labels provided in the PianoSyllabus website metadata.} Attending to Fig.~\ref{fig:era_distro}, the Romantic era leads with the highest number of pieces, closely followed by the \review{western classical music of the 20th Century}, each \review{amounting to} more than 2\,000 pieces. The \review{Modern period, which primarily includes popular genres such as rock,} exhibits a slightly lower count yet maintains a remarkable presence in the dataset. In contrast, the Baroque and Classical styles \review{feature fewer pieces}, indicating a larger concentration of works in the later musical periods within this particular dataset.

\begin{figure}[ht!]
\centering
\includegraphics[trim=40pt 10pt 50pt 20pt, clip,width=.95\linewidth]{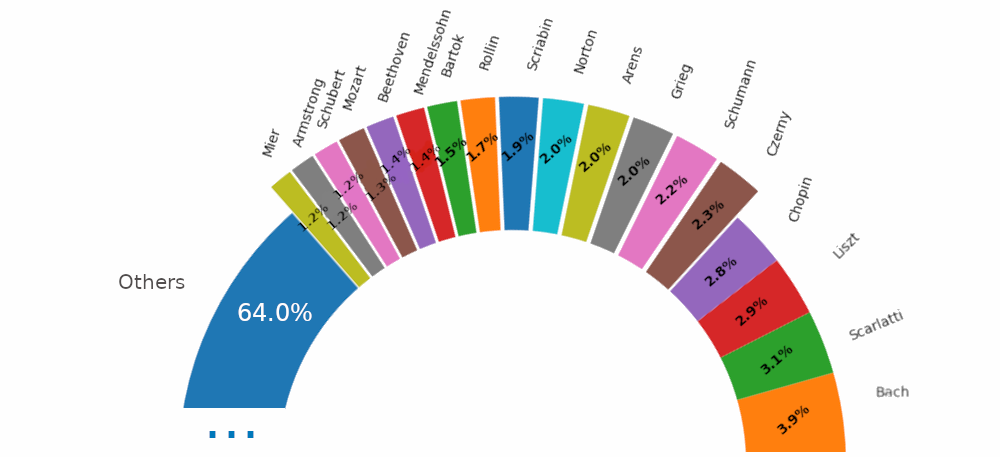}
\caption{\review{Composer} distribution of PSyllabus dataset.}
\label{fig:compo_distro}
\end{figure}

The PSyllabus, dataset comprises a large number of composers, whose \review{exact distribution over} the assortment is presented in Fig.~\ref{fig:compo_distro}. As it may be observed, a \review{64.0\%} of the collections corresponds to the `Others' tag, indicating a long tail of composers with smaller representation in the dataset \review{that results in an average number of 4 tracks per composer}. The remaining portion is divided among several well-known composers, with the largest represented groups being D.~Scarlatti, F.~Liszt, J.~S.~Bach, and F.~Chopin, each constituting \review{close to a 3.0\%} of the dataset. Following them, a gradual decrease in representation is observed with composers like E.~Grieg, C.~Czerny, and others, \review{each one representing, in the largest case, a 2\% of the dataset}.  This suggests a diverse dataset with a few composers being more frequently represented than a vast number of others.

\begin{figure}[ht!]
\centering
\includegraphics[trim=5pt 10pt 5pt 5pt, clip,width=.95\linewidth]{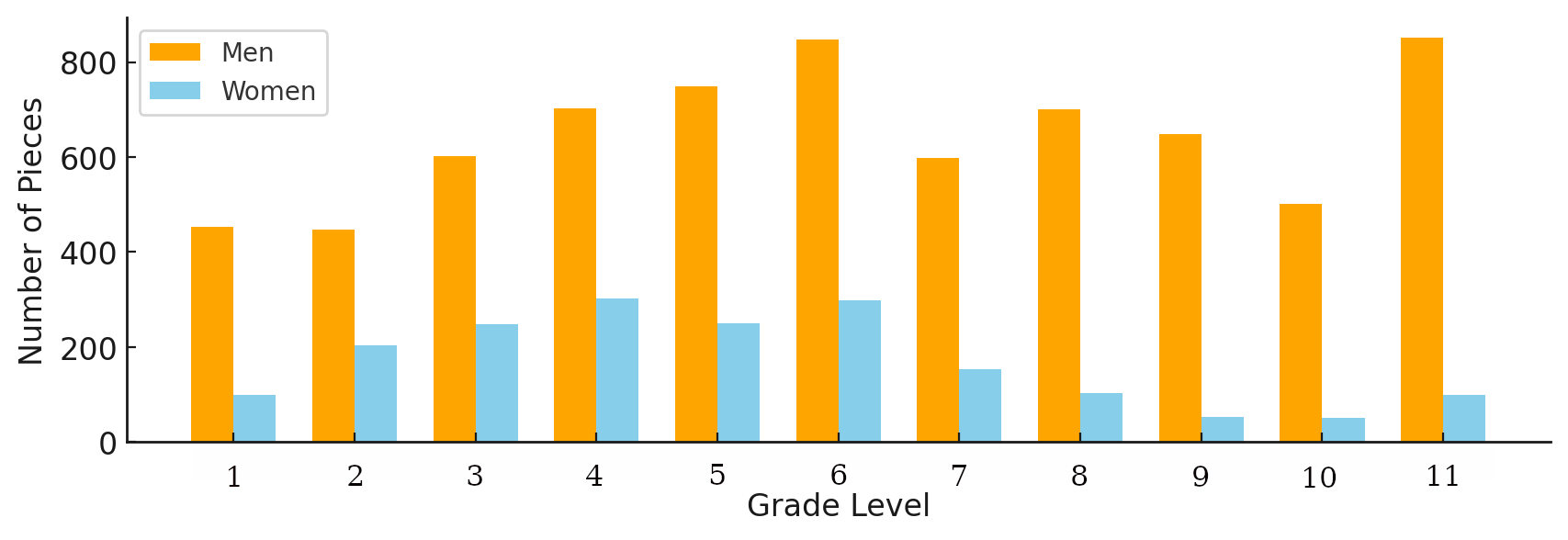}
\caption{Distribution of the pieces of the PSyllabus dataset among the different difficulty levels. Gender distribution is also provided for the sake of analysis.}
\label{fig:woman_men}
\end{figure}

Regarding the difficulty levels of the PSyllabus dataset, Fig.~\ref{fig:woman_men} displays the \review{distribution of pieces} among the different grades and gender of the composer. A first point that may be highlighted out of the provided graph is that works of male composers spread across all the grades, with the highest number of pieces at level 10 and a notable presence in \review{levels 2 to 5}. The contributions of female composers, on the other hand, concentrate around grade 3. Although the level distribution of the women peaks at a lower grade level compared to men, the PSyllabus dataset includes over 14\% of pieces by female composers, \review{a high representation of women composers specifically within the context of the difficulty estimation task~\cite{ramoneda2024,ramoneda2023predicting}}. This is a positive step towards promoting the repertoire of a group that has been historically underrepresented, yet it underscores the ongoing need for efforts towards the support and enhancement of the visibility of women composers in the musical canon.

Finally, to gather additional insights about the consistency of the difficulty assignments in the PSyllabus dataset, we compare the annotations with those in other well-known rankings provided by established examination boards: Australian Guild of Music Education (AGME), Trinity College London (Trinity), Australian Music Examinations Board (AMEB), Royal Conservatory of Music (RCM), General Certificate of Secondary Education (GCSE), St. Cecilia School of Music (SCSM), Royal Irish Academy of Music (RIAM), Associated Board of the Royal Schools of Music (ABRSM), New Zealand Music Examinations Board (NZMEB), Australian and New Zealand Cultural Arts (ANZCA), Performance Series (PS), London College of Music (LCM), and the web community Piano Street (Piano St).

\begin{figure}[bt!]
\centering
\includegraphics[width=.85\columnwidth]{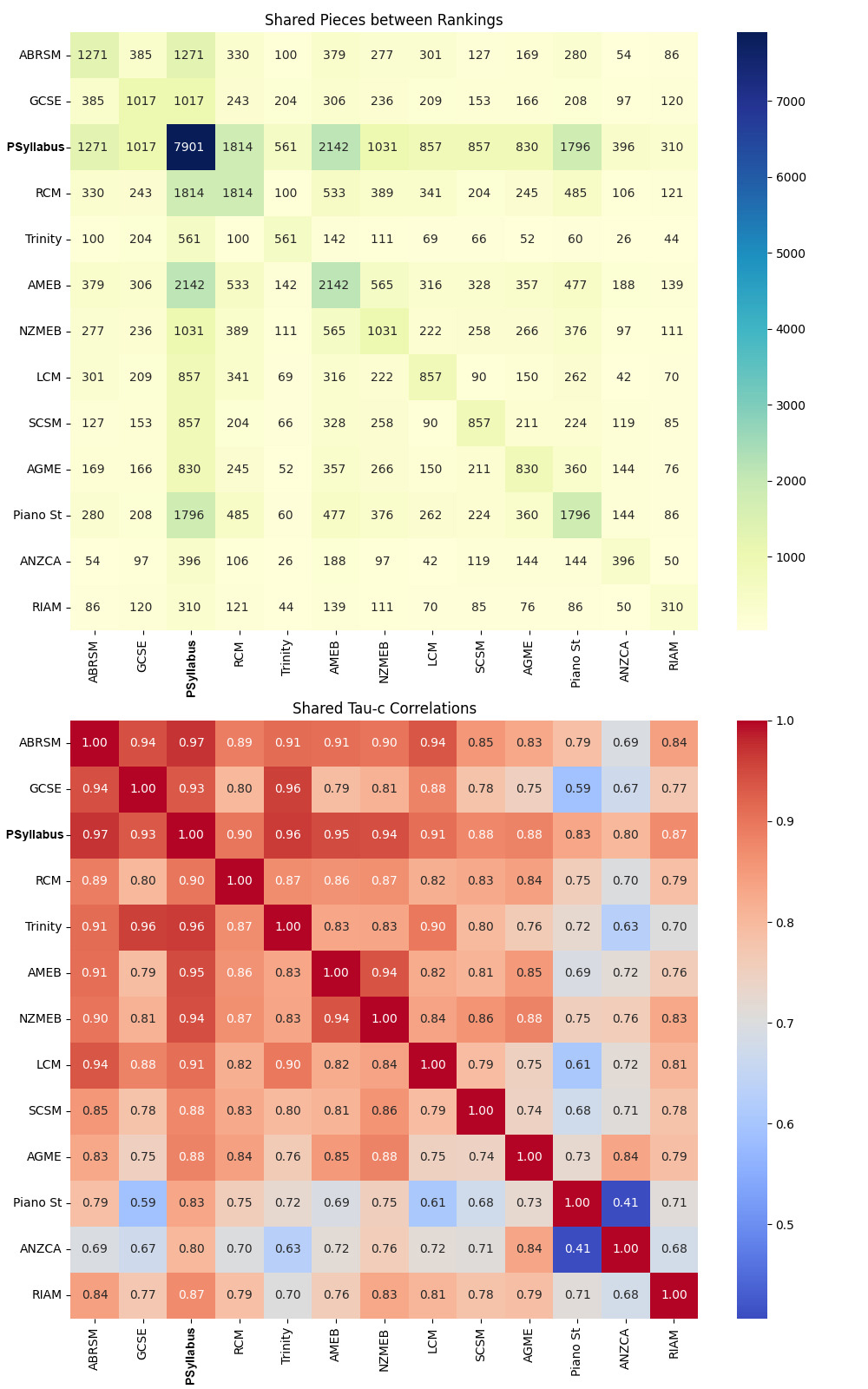}
\caption{Shared difficulty pieces and their ranking correlation between PSyllabus and other rankings.}
\label{fig:correlations}
\vspace{-0.5cm}
\end{figure}

We resort to the Kendall rank correlation coefficient ($\tau_c$) to perform these comparisons. This metric, which performs a pair-wise assessment of two ranks by attending to the relative order among their elements, represents a robust procedure to analyze rank correlations. \review{It is important to note that the analysis is limited to the pieces included in the PSyllabus dataset to ensure consistency across comparisons.} Based on this premise, Fig.~\ref{fig:correlations} provides the results of this correlation analysis. In addition, since the different rankings do not \review{account for} the same music pieces, this figure also provides the absolute number of works shared for each pair of rankings.

One of the highest \review{similarity} scores ($\tau_c=0.97$) is observed when comparing ABRSM with PSyllabus, depicting that relative difficulty order in their pieces hardly differs between both rankings. On average, correlation across all the rankings is $\tau_c=0.81$ (with a standard deviation of 0.099), which suggests that most schemes agree on the \review{ranking of pieces with} slight differences among them. \review{It may be also observed that all correlations between the PSyllabus and the rest of the rankings surpass the commented average correlation score, hence proving the consistency in the annotations.}

\subsection{Benchmark datasets}
\label{benchmark_datasets}


This section describes the two additional benchmark datasets presented in \review{this} work: the Hidden Voices dataset of underrepresented composers and the PSyllabus multiple performances collection. These assortments are now described:

\subsubsection{Hidden Voices dataset}
This first dataset comprises pieces of the \review{Hidden Voices Project~\cite{hidden} annotations and audio music recordings from YouTube}. The result is a collection of 57 compositions with 7 difficulty levels: \textit{elementary} (1 piece), \textit{late elementary}  (1 piece), \textit{early intermediate} (3 pieces), \textit{mid intermediate} (17 pieces), \textit{late intermediate} (11 pieces), \textit{early advanced} (3 pieces), and \textit{advanced} (14 pieces).

\subsubsection{PSyllabus multiple performances collection}
The rationale behind this benchmark is that of collecting multiple performances for certain pieces that differ in specific conditions such as being played by different musicians\review{~\cite{zhang2022atepp}}, piano models, various tempi, etc. For each composition, we collected \review{different interpretations} from YouTube assuring that none of the groups \review{contained duplicated recordings of the same performance}. Moreover, different segments from a single YouTube resource were treated as different performances for videos that include multiple performances (e.g., tutorial videos). Eventually, this set provides 55 pieces, which represents 5 different interpretations for 11 distinct piano pieces---one per difficulty level in PSyllabys---from 5 music eras.

\section{Methodology}
\label{methodology}
Difficulty estimation is typically addressed as an \textit{ordinal classification task} (also known as \textit{ordinal regression})~\cite{ramoneda2023predicting,ramoneda2024}. Such a recognition framework adequately models the task at hand since the difficulty levels do not constitute unrelated and isolated categories but they represent an ordered set. This formulation \review{poses certain considerations} compared to general classification tasks that are later discussed. 

Formally, let $\mathcal{X}$ be an input representation space---i.e, the acoustic recording---and $\mathcal{Y}$ denote the target difficulty space, related by the underlying function $f:\mathcal{X}\rightarrow\mathcal{Y}$. \review{The goal approximate $f$ using a set of annotated data $\mathcal{T}\subset\mathcal{X}\times\mathcal{Y}$.}

Considering this framework, the target space of $C$ difficulty levels for a given datum $x_i\in\mathcal{X}$ is represented as a mutilabel vector $\mathbf{y}_i = \left[\lambda_1,\ldots,\lambda_C\right]$ where $\lambda_j\in\left\{0,1\right\}$ with $1\leq j\leq C$ and $\lambda_j \leq \lambda_{j-1}\;\forall\;j > 1$. \review{That is, this encoding ensures that element $\lambda_j$ may only present a value of 1 if and only if all its preceding elements in the multilabel vector $\mathbf{y}_i$---i.e., elements in positions $1$ to $j-1$ in vector $\mathbf{y}_i$---also display a value of 1. This condition forces a monotonic relation between the $\lambda_j$ elements of multilabel vector $\mathbf{y}_i$ imposed by the ordinal recognition formulation considered~\cite{cheng2008neural}.}

\review{Note that, the $\mathbf{y}_i$ multilabel vector is mapped onto the class label that represents the difficulty level using function $\zeta:\left\{0,1\right\}^{C} \rightarrow \mathbb{N}\cap \left[1, C\right]$ defined as:}
\review{
\begin{equation}
    \zeta\left(\mathbf{y}_{i}\right) = \underset{l\in\mathbb{N}\cap\left[1,C\right]}{\operatorname{argmax}}\left(\left|\lambda_{j}=1\;\forall\;j\leq l\right| = l\right)
\end{equation}
}

\review{The remainder of the section presents the details related to the input representation $\mathcal{X}$ (Section~\ref{ssec:input}), the unimodal recognition model (Section~\ref{ssec:frontend}), and its multimodal version (Section~\ref{ssec:multimodal}).}

\subsection{Input representations}
\label{ssec:input}
Existing approaches for performance difficulty estimation in the literature have exclusively focused on the cases of either symbolic data or sheet music images. Hence, finding the most suitable input representation remains an open question, for which we compare and assess two possibilities: 
\begin{enumerate}
    \item{\textbf{Spectral \review{mid-level} representation:}} We resort to the Constant-Q Transform (CQT)~\cite{schorkhuber2010constant} \review{due to its reported benefits in different \review{MIR} tasks~\cite{cheuk2020impact} as well as to the fact that it suits the equal-temperament tuning commonly used in piano music. While other representations (e.g., mel-spectrograms) were initially studied, they were eventually discarded in favor of the CQT one \review{based on} its performance superiority shown on preliminary experimentation.}

    \item{\textbf{Piano roll representation:}} \review{Based on} its success in various MIR tasks~\cite{kong2020giant}, we consider the transcription framework by Kong et al.~\cite{transcription2022kong} \review{pretrained on Maestro dataset~\cite{hawthorne2018enabling}}, which retrieves two matrices: (i) the frame-wise pitch activations, which naturally embed the offset information; and (ii) the onset information, which is explicitly provided to account for repetitive onsets without offsets (e.g., when using the sustain pedal). \review{We clarify that the piano roll information is constructed from the transcribed MIDI data, rather than directly from the raw activation output of the model.}

\end{enumerate}

\review{Both input representations are formally denoted in the rest of the work as} $\mathcal{X}\subset\mathbb{R}^{b\times t\times c}$, where $b$ and $t$ respectively denote the number of bins and time \review{steps for} the signal, while $c$ stands for the amount of channels ($c=1$ for CQT and $c=2$ for piano roll). \review{Finally, the number of bins is set to $b=88$, which ensures higher frequency resolution necessary for piano transcription, especially for high notes. For other MIR tasks, CQT bins for each semitone are commonly used, as discussed in Krause et al.~\cite{krause2021leitmotif}.}

\subsection{Recognition framework}
\label{ssec:frontend}

Due to its strong performance in the MIR literature, we use a Convolutional-Recurrent Neural Network (CRNN) to estimate difficulty in piano recordings. Unlike other MIR classification models, this framework handles variable-length input signals, which is essential to avoid constraining recording duration. The following section details the model, as shown in Fig.~\ref{fig:attention_gru}.

\begin{figure}[h!]
  \centering
  \includegraphics[width=.7\columnwidth]{imgs/main_arquitectura.jpg}
    \caption{Recognition framework for piano performance difficulty estimation. The input signal is analyzed with a residual convolutional network for feature extraction, followed by a recurrent stage with attention mechanisms to summarize the information that is later processed by the classification stage.}
    \label{fig:attention_gru}
\end{figure}

Given a datum $x\in\mathbb{R}^{b\times t\times c}$, the \review{model comprises} an initial convolutional network with residual connections that retrieves a feature map $x^f\in\mathbb{R}^{b^\prime \times t^\prime \times c^\prime}$, which is then reshaped into sequence $\mathbf{F} = \left[F_{t^\prime_0}, \ldots, F_{t^\prime_n}\right]\in\mathbb{R}^{t^\prime \times (b^\prime \cdot c^\prime)}$ of features. This sequence is then processed by the recurrent stage that retrieves \review{$\mathbf{Z} = \left[Z_{t^\prime_0}, \ldots, Z_{t^\prime_n}\right]\in\mathbb{R}^{t^\prime \times m}$}\review{, where $m\in\mathbb{N}$ denotes the embedding size}, to model the temporal dependencies among the computed features maps.

After that, based on its success in areas involving sequential data such as text summarization~\cite{yang2016hierarchical} or score analysis~\cite{jeong2019virtuosonet}, the model incorporates a \review{\textit{context attention}} stage to summarize the $\mathbf{Z}$ sequence into a single vector \review{$R\in\mathbb{R}^{m}$} \review{based on attention mechanisms}. This representation is then fed to a neural-based linear classifier that retrieves the $C$-sized multilabel vector $\hat{\mathbf{y}}$ depicting the estimated difficulty of the piece.

During the inference stage, \review{a threshold is applied each element in vector $\hat{\mathbf{y}}$ so that values above and below are respectively converted into 1 and 0.} This value, which in this work has been fixed to $0.5$, constitutes a hyper-parameter to be adjusted by the user.

Finally, it must be highlighted that the presented model describes the case in which the training procedure only considers a single task. In the event that \review{several of them are concurrently employed}, we must include as many classification/regression heads as the number of tasks considered.

\subsection{Multimodal approach}\label{ssec:multimodal}

Based on the premise that the two input representations may complement each other in certain aspects, we consider a multimodal framework with the aim of leveraging their individual strengths. More precisely, our hypothesis is that \review{the mid-level and transcription representations may complement each other in a synergic manner that could eventually boost the recognition performance.}

To do so, we propose the early-fusion scheme, denoted as \textit{MM} in the rest of the work, which is now detailed. Given a piano recording to be processed, we initially derive the two input representations \review{used} in the work, i.e., the \review{CQT} $x^{cqt}\in\mathbb{R}^{b\times t \times 1}$ and the piano-roll transcription $x^{pr}\in\mathbb{R}^{b\times t \times 2}$. Once these representations have been obtained, they are concatenated at the channel level, retrieving the multimodal encoding $x^{mm}\in\mathbb{R}^{b\times t \times 3}$, which is processed by the recognition framework in Section~\ref{ssec:frontend}.


It must be highlighted that this early-fusion scheme depicts the inherent limitation that the number of bins $b$ as well as the time steps $t$ must match in both input representations. \review{In this regard, the analysis parameters in both representations, which are reported in Section~\ref{ssec:input}, are deliberately tuned so that this alignment is guaranteed.}

Finally, based on the reported competitive performance of late-fusion multimodal schemes in different tasks in MIR~\cite{alfaro2022late}, we consider an ensemble-based classification approach following symbolic difficulty estimation \cite{ramoneda2024} for comparative purposes. This strategy, denoted as ENSEMBLE in the rest of the work, involves using a deep learning ensemble classifier to enhance overall performance by combining multiple models. Specifically, this strategy involves training one recognition model for each possible input representation---i.e., the \review{CQT} and the transcribed piano roll---that, in inference time, will retrieve one individual estimation per modality, respectively denoted as \( \hat{\mathbf{y}}_{\scriptscriptstyle \text{CQT}} \) and \( \hat{\mathbf{y}}_{\scriptscriptstyle \text{PR}} \). These predictions are eventually averaged to make the final decision, i.e., \( \hat{\mathbf{y}}_{\scriptscriptstyle \text{ENSEMBLE}} = \frac{\hat{\mathbf{y}}_{\scriptscriptstyle \text{CQT}} + \hat{\mathbf{y}}_{\scriptscriptstyle \text{PR}}}{2} \). Note that, as in the early-fusion proposal, this approach is expected to improve the results compared to the base unimodal strategies by leveraging the strengths of each individual modality and representation.




\section{Experimental setup}\label{sec:experimental_setup}
This section presents the details related to the datasets used in the work as well as the signal representations used, the evaluation metrics to assess \review{our hypothesis}, and the details related to the training procedure of the recognition scheme.

\subsection{Data collections}

We consider the PSyllabus dataset presented in \review{this} work for both training and evaluating the recognition model. For that, we resorted to a 5-fold cross-validation scheme, devoting the 60\% of \review{the} dataset as train partition and the rest as equally-sized validation and test sets. For the sake of analysis, we also evaluated the performance of the model on the two benchmark datasets introduced in the work---i.e., Hidden Voice and PSyllabus with multiple performances---, being these assortments exclusively used for testing purposes.

Regarding the input acoustic signal, we consider two different representations: (i) a raw time-frequency one based on the \review{CQT} with a hop length of 160 samples, a total number of 88 bins, and 12 bins per octave; (ii) and a piano roll representation obtained with the AMT method by Kong et al.~\cite{transcription2022kong}, fixing the output space to 88 notes to match that of the CQT.

Once the representations are obtained, we postprocess them to reduce their \review{size} in the temporal dimension by subsampling to 5 frames per second. \review{This choice ensures operational efficiency and allows full audio sequences to be processed within the memory limits. Initial experiments showed this frame rate provides sufficient accuracy for difficulty prediction. While higher frame rates or alternative methods (e.g., strides or max-pooling) could improve temporal detail, they would require resources beyond our current setup.} However, we acknowledge exploring higher frame rates could improve model performance by providing more detailed temporal data. 

Finally, it must be highlighted that, besides the difficulty ranking annotations in the PSyllabus dataset, we also consider its annotations related to the music era/period, the composer, and other ranking labels for the multi-task training framework.

\subsection{Performance evaluation}

Regarding performance evaluation, we utilized two commonly employed metrics in ordinal classification~\cite{gaudette2009evaluation} previously considered in the literature of performance difficulty estimation~\cite{ramoneda2023predicting}: \review{\textit{classification accuracy} ($\mbox{Acc}$) and \textit{mean squared error} (MSE)}. \review{Note that, the former figure of merit frames the task within a classification paradigm, whereas the latter evaluates the model in terms of a regression task.}




 \review{Finally}, we consider the $\tau_c$ correlation coefficient to evaluate the correctness of the estimated rankings in comparison to the ground-truth one. This statistical method is particularly relevant to our study: as outlined in Section~\ref{dataset} in the analysis of Fig.~\ref{fig:correlations}, an average agreement of $\tau_c=0.81$ is reported among expert human \review{evaluators} on the task; in this regard, it is of relevance \review{to} assess the proposed method in these terms to comparatively evaluate its predictive capabilities against those of expert human evaluators. 

\subsection{Training procedure}

The architecture discussed in Section~\ref{ssec:frontend} features a recognition framework integrating CNNs, GRUs, and an attention mechanism for audio data analysis. The CNN component employs three residual blocks, each with 3x3 convolutional layers, batch normalization, and ReLU activations, interconnected with max-pooling and dropout layers. The piano roll and CQT models use these blocks with output channels of 64, 128, and 256. In the multimodal configuration, separate CNN branches for piano roll and CQT inputs are concatenated. The GRU layer consists of two bidirectional layers with an input size of 256 (512 for multimodal inputs) and a hidden size of 128. A context attention mechanism with four heads refines the features by computing similarity with a learned context vector and applying softmax for attention weights. The classifier comprises a fully connected layer with 256 input units, followed by a ReLU activation and an output layer for classification. \review{We conducted initial experimentation to set the hyperparameters due to limited resources, instead of employing exhaustive grid search or other optimization techniques. We acknowledge that better hyperparameter configurations may exist and could be explored in future work to improve performance.}

\review{The recognition model was trained on a RTX 2080 GPU using a mean squared error loss as in the work by Cheng et al.~\cite{cheng2008neural}}\footnote{\review{As commented by Cheng et al.~\cite{cheng2008neural}, the ordinal information is embedded in the task by the multilabel encoding presented in Section~\ref{methodology}, being hence possible to train the scheme with any loss function that suits the case.}} with a batch size of 16 samples. \review{The training process uses complete pieces as input to ensure the preservation of long-term contextual dependencies in the data.} The Adaptive Moment Estimation (Adam) optimizer~\cite{jin2016deep} was used with a learning rate of $10^{-3}$. A gradient clipping threshold of $10^{-4}$ with \review{a weight decay policy} were included to ensure stable training conditions. To prevent overfitting, we incorporated an early-stopping mechanism that monitors both the \review{Acc} and MSE metrics of the validation set. \review{The early-stopping mechanism used a patience of 50 epochs, halting the training process if no improvement was observed in either metric within this window.}

The previously explained training procedure refers to the single-task case---i.e., only using the difficulty level to train the model---, the multi-task setting maintains this scheme and \review{adds} additional loss terms for the other tasks: on the one hand, the composer recognition case considers the categorical cross-entropy loss; on the other hand, for the period/era estimation as well as the additional difficulty rank information, we resort to the ordinal loss of the single-task scenario.

\section{Results}
\label{sec:results}

This section analyzes the results obtained considering the aforementioned experimental scheme. For the sake of clarity, Section~\ref{sec:res_input_representations} \review{assesses} the influence of the input representation on the overall performance of the scheme whereas Section~\ref{sec:res_auxiliary_tasks} \review{explores} the use of auxiliary tasks for improving the recognition capabilities of the method. \review{Note that, for the sake of clarity, arrows $\uparrow$ and $\downarrow$ respectively denote whether the reported metric is optimized as it is increased or decreased.} \review{Additionally, bold values in the tables indicate the best results within each experimental setup. The value without parentheses represents the mean of five experiments conducted across different train, validation, and test splits, while the value in parentheses corresponds to the standard deviation.}

\subsection{Input representation analysis}\label{sec:res_input_representations}

The first experiment of the work analyses the influence of input representation on the overall performance of the scheme. For that, we assess the performance of the \review{CQT} and piano roll (PR) representations in both a unimodal manner as well as in a multimodal one, considering the early and late fusion procedures---respectively denoted as MM and ENSEMBLE---described in Section~\ref{ssec:multimodal}. The results of these experiments are provided in Table~\ref{tab:res_input_representations}.


\begin{table}[h!]
\vspace{-0.2cm}
\centering
\caption{Results training with unimodal representations, CQT and PR, and multimodal ones.}
\resizebox{.9\columnwidth}{!}{%
\begin{tabular}{llccc}
\toprule[1pt]
\multicolumn{2}{l}{\textbf{Experiment}}  &    \textbf{MSE} ($\downarrow$)    & \review{\textbf{Acc}} ($\uparrow$)    & $\mathbf{\tau_c}$ ($\uparrow$)     \\ \hline
\multicolumn{2}{l}{\textit{Unimodal}}\\
 &CQT & 2.29 (.18) & 32.9 (1.6) & .74 (.01)\\
 &PR  & \textbf{1.85 (.07)} & \textbf{36.7 (.8)} & \textbf{.77 (.01)}\\ 
\hdashline
\multicolumn{2}{l}{\textit{Multimodal}}\\
&MM & \textbf{1.81 (.11)} & \textbf{37.3 (2.0)} & \textbf{.78 (.01)}\\ 
&ENSEMBLE & 1.82 (.09) & 36.2 (.6) & .78 (.01)\\ 
\bottomrule[1pt]
\end{tabular}
}
\label{tab:res_input_representations}
\end{table}

Attending to the results obtained for unimodal scenarios, PR consistently outperforms the CQT representation for all evaluation metrics. This fact suggests that a symbolic-level representation---i.e., PR---may be more appropriate in identifying crucial patterns for this difficulty estimation task rather than a raw time-frequency representation---i.e. CQT. 

Regarding multimodal scenarios, it may be observed that the MM scheme depicts the best recognition performance of the studied cases. Such a result proves the superiority of the early fusion of the input representations rather than the late combination of the individual decisions, possibly due to the fact that the model is able to learn an adequate representation for each modality that it later appropriately \review{exploited}.

\review{Given the narrow improvement observed in the multimodal scheme with respect to the unimodal approaches---and, especially, with the PR representation---the rest of the section exclusively focuses on these latter cases to better understand the performance gap between them.}

\review{Finally, while not directly comparable to other works from the literature due to the different nature of the data~\cite{ramoneda2023predicting,ramoneda2024}, the results in this work prove the higher challenge that the audio modality poses on the task since the figures obtained are considerable lower than those of the reference works.}

\subsection{Multi-task training schemes}\label{sec:res_auxiliary_tasks}

Having analyzed the influence of the input representation on the success of the task, we now explore the use of multi-task training with the same goal. For that, considering the base recognition model trained exclusively on the difficulty labels of the PSyllabus collection, we incorporate a set of auxiliary tasks: composer identification, musical era/period identification, and \review{additional difficulty rankings derived from external sources or alternative annotations. These ``rank'' annotations serve as auxiliary tasks to complement the main difficulty estimation objective by providing additional context about the relative difficulty of the pieces.} The results obtained are provided in Table~\ref{tab:res_auxiliary_tasks}.


\begin{table}[h!]
\setlength{\tabcolsep}{3.0pt}
\renewcommand{\arraystretch}{1.1}
\centering
\caption{Result of the single- and multi-task training schemes for the different input representations considered.}
\resizebox{\columnwidth}{!}{%
\begin{tabular}{lccccc}
\toprule[1pt]
\multirow{2}{*}{\textbf{Metric}} & \multirow{2}{*}{\textbf{Representation}} & \multirow{2}{*}{\textbf{Single}} & \multicolumn{3}{c}{\textbf{Multi-task}}\\
\cmidrule(lr){4-6}
 &  &  & \textbf{Era} & \textbf{Composer} & \textbf{Ranks}\\
\cmidrule(r){1-2}\cmidrule(l){3-6}
\multirow{2}{*}{MSE ($\downarrow$)}  & CQT & 2.29 (.18) & \textbf{1.99 (.15)} & 2.74 (.16) & 2.16 (.16)\\
 & PR & 1.85 (.07) & \textbf{1.83 (.10)} & 2.57 (.18) & 1.88 (.06)\\
 \cmidrule(lr){1-6}
\multirow{2}{*}{\review{Acc} ($\uparrow$)} & CQT &  32.9 (1.6) & \textbf{35.6 (1.0)} & 27.0 (1.1) & 34.4 (1.4)\\
 & PR & 36.7 (.8) & \textbf{37.7 (.9)} & 29.0 (1.1) & 35.4 (.6)\\
 \cmidrule(lr){1-6}
\multirow{2}{*}{$\tau_c$ ($\uparrow$)} & CQT & .74 (.01) & \textbf{.77 (.01)} & .70 (.01) & \textbf{.75 (.01)}\\
 & PR & \textbf{.77 (.01)} & \textbf{.78 (.01)} & .71 (.01) & \textbf{.77 (.01)}\\
\bottomrule[1pt]
\end{tabular}
}
\label{tab:res_auxiliary_tasks}
\end{table}

The integration of multi-task learning into the training of the recognition scheme resulted in \review{specific trends of improvement in performance metrics, rather than uniform enhancements across all tasks}. This \review{trend} is particularly evident in the \textit{era classification} task, which consistently achieves the best performance results across all training scenarios, including both single-task and multi-task cases.

On the contrary, the \textit{composer identification} task does not report any benefit to the recognition scheme as it not only reports the lowest performance among the multi-task scenarios, but also underperforms the single-taverleafsk case for all evaluation metrics. This is possibly due to the fact that the diverse stylistic idiosyncrasies of each composers might impede the generalization capabilities of the model.

Experiments involving the \textit{multiple ranking} task yielded mixed outcomes. Although there was a slight improvement compared to the single-task baseline, these enhancements were not as pronounced as those observed in the \textit{era identification} task. It must be noted that these outcomes contradict those involving music sheet images in previous research~\cite{ramoneda2023predicting}, where the multi-rank approach outperformed other scenarios. However, a direct comparison between these results is not straightforward due to differences in the datasets used, and further research is required to adequately validate these outcomes.

\subsubsection{Impact of the musical era on the difficulty estimation}
Based on the reported success of the \textit{era identification} task within the multi-task training scenario, this section further analyses these results to gain additional insights on it. More precisely, the idea in this case is assessing whether the improvement observed may be due to biases introduced by the music period. Based on this premise, Table~\ref{tab:periods_experiments} shows the results obtained when considering the \textit{era identification} task of the multi-task approach for each particular music period.

\begin{table*}[ht!]
\vspace{-.5cm}
\centering
\caption{Comparative analysis for Basic and Multi-task with Era experiments across musical periods.}
\label{tab:periods_experiments}
\resizebox{.8\textwidth}{!}{%
\begin{tabular}{llcccccc}
\toprule[1pt]
\multirow{2}{*}{\textbf{Method}} & \multirow{2}{*}{\textbf{Period}} & \multicolumn{2}{c}{\textbf{MSE} ($\downarrow$)} & \multicolumn{2}{c}{\review{\textbf{Acc}} ($\uparrow$)} & \multicolumn{2}{c}{$\mathbf{\tau_c}$ ($\uparrow$)} \\
\cmidrule(lr){3-4} \cmidrule(lr){5-6} \cmidrule(lr){7-8}
       &        & \textbf{Single} & \textbf{with Era} & \textbf{Single} & \textbf{with Era} & \textbf{Single} & \textbf{with Era} \\
\cmidrule(lr){1-8}
\multirow{5}{*}{CQT}
& Baroque & 2.73 (.49) & 2.49 (.42) & 28.1 (4.8) & 32.3 (5.5) & .62 (.04) & .63 (.04) \\
& Classical & 2.25 (.66) & \textbf{1.96 (.51)} & 32.2 (6.1) & 34.1 (5.6) & .70 (.09) & .72 (.07) \\
& Romantic & 2.16 (.56) & 2.05 (.44) & 32.6 (5.4) & 34.1 (4.5) & .70 (.07) & .72 (.06) \\
& 20\textsuperscript{th} Century & 2.16 (.49) & 2.33 (.42) & 33.3 (4.9) & 3.3 (6.6) & .72 (.07) & .66 (.04) \\
& Modern & 2.25 (.50) & \textbf{1.84 (.47)} & 32.9 (4.7) & \textbf{37.0 (6.1)} & .70 (.07) & \textbf{.75 (.06)} \\
\hdashline
\multirow{5}{*}{PR}
& Baroque & 2.25 (.50) & \textbf{2.03 (.62)} & 32.9 (4.7) & \textbf{34.2 (6.5)} & .70 (.07) & \textbf{.71 (.09)} \\
& Classical & 1.97 (.79) & 1.98 (.45) & 34.2 (8.2) & \textbf{34.3 (4.9)} & .73 (.09) & \textbf{.73 (.07)} \\
& Romantic & \textbf{1.84 (.66)} & 1.87 (.63) & 35.2 (6.9) & \textbf{35.4 (7.6)} & \textbf{.74 (.08)} & .73 (.09) \\
& 20\textsuperscript{th} Century & 1.86 (.57) & \textbf{1.78 (.52)} & 35.6 (6.1) & \textbf{37.3 (7.0)} & \textbf{.75 (.07)} & .74 (.07) \\
& Modern & 1.94 (.57) & 1.92 (.50) & 35.8 (5.5) & 36.3 (5.6) & .74 (.06) & .74 (.06) \\
\bottomrule[1pt]
\end{tabular}
}
\vspace{-.3cm}
\end{table*}

As it may be observed, for the Baroque period the MSE figure improves from 2.73 to 2.49 for the CQT representation and from 2.25 to 1.84 for the PR one, with corresponding increases in the \review{Acc} metric. In a more moderate manner, the Classical, Romantic, and Modern periods depict similar trends, demonstrating the general efficacy of integrating era information into difficulty estimation models.

These findings highlight the significance of era-specific characteristics in music difficulty prediction, suggesting a nuanced relationship between historical context and model accuracy. \review{The varied performance improvements across periods likely stem from the distinct musical characteristics inherent to each era, such as harmonic complexity, stylistic conventions, or common compositional techniques. These traits may affect how well the model can generalize predictions within a specific period. This observation underscores the potential of developing tailored, era-specific modeling approaches to further enhance prediction accuracy.}

\section{A case study on woman composers}
\label{sec:case_women}
This section proposes a case study on the classification of women composers---an \review{underepresented} group in general music repertoires---using the proposed recognition model. For that, we perform two experiments to further gain insights about the recognition capabilities of the framework: a first one in which we analyze its gender bias on the PSyllabus dataset (Section~\ref{ss:gender_gap}) and a second one in which we assess its generalization capabilities using the Hidden Voices benchmark dataset that exclusively comprises black women composers (Section~\ref{ss:zero_shot}). The models used in these experiments are presented in Table~\ref{tab:res_input_representations}, and only used for inference. 

\subsection{Model performance on works by female composers}
\label{ss:gender_gap}

To analyze the gender bias, we assess the recognition performance of the scheme when targeting pieces created by composer belong to a specific gender, i.e., either male or female composers. Table~\ref{tab:gender_gap} presents the recognition results obtained for this experiment considering the PSyllabus dataset.

\begin{table}[h!]
\centering
\caption{Analysis of model performance differentiated by the composer's gender.}
\resizebox{.9\columnwidth}{!}{%
\begin{tabular}{lcccc}
\toprule[1pt]
\multicolumn{1}{l}{\textbf{Method}} & \textbf{Experiment}  &    \textbf{MSE} ($\downarrow$)     & \review{\textbf{Acc}} ($\uparrow$)    & $\mathbf{\tau_c}$  ($\uparrow$)     \\
\hline
\multirow{3}{*}{CQT} & \textit{Both genders} & 2.29 (.18) & 32.9 (1.6) & .74 (.01)\\
& \textit{Only men} & 2.24 (.19) & 33.2 (1.1) & .73 (.01)\\
& \textit{Only women} & 2.47 (.59) & 31.6 (4.7) & .67 (.03)\\
\hdashline
\multirow{3}{*}{PR} & \textit{Both genders} & 1.85 (.07) & 36.7 (.8) & \textbf{.77 (.01)}\\
& \textit{Only men} & 1.83 (.05) & 37.3 (.8) & .76 (.01)\\
& \textit{Only women} & 1.91 (.60) & 34.6 (4.2) & .72 (.02)\\
\hdashline
\multirow{3}{*}{MM} & \textit{Both genders} & \textbf{1.81 (.11)} & \textbf{37.3 (2.0)} & \textbf{.78 (.01)}\\
& \textit{Only men} & \textbf{1.81 (.09)} & \textbf{37.7 (2.5)} & \textbf{.77 (.01)}\\
& \textit{Only women} & \textbf{1.72 (.49)} & \textbf{35.7 (5.7)} & \textbf{.73 (.03)}\\
\bottomrule[1pt]
\end{tabular}
}
\vspace{-0.1cm}
\label{tab:gender_gap}
\end{table}

The analysis reveals that, despite the uniform training performed across all models, recognition disparities emerge when testing on gender-specific data. The MM model, when tested on mixed-gender data, exhibited the best performance with an accuracy of Acc$_0=37.3\%$ and a $\tau_c=0.78$, indicating a well-rounded capability to interpret compositions by both genders. However, a remarkable performance decrease was observed for the set of women composers. \review{While the mean MSE was lower for this group, the standard deviations of both MSE and Accuracy were significantly higher compared to the male composer group. This suggests that, at a coarse level, the model predictions for women composers align closely with the ground truth (low MSE), but at a finer level, they deviate from the expected trends (low Accuracy). We hypothesize that this discrepancy may stem from inherent differences in musical characteristics or potential biases in the difficulty annotations for compositions by women composers.} These results underscore the importance of diverse and balanced evaluation sets to avoid model biases and ensure fair performance. Further research is needed to better understand these performance differences and determine their origin, a question currently being explored in collaboration with a musicology researcher.

\subsection{Zero-shot experiment on Hidden Voices collection}
\label{ss:zero_shot}

This second experimental scenario related to women composers focuses on the performance assessment of the model in a \review{zero-shot scenario}. Specifically, the model was trained on the PSyllabus dataset, which uses an ordinal difficulty system with 11 labels, and evaluated on the Hidden Voices benchmark, which employs a different ordinal system with 7 labels and distinct boundaries. \review{The model had no exposure to the Hidden Voices data during training, making this a true zero-shot setting that evaluates its ability to generalize across datasets with differing labeling schemes.}

Table~\ref{tab:zero} shows the results obtained \review{regarding} the $\tau_c$ correlation coefficient, considering both the single-task and multi-task models posed, as well as the two considered input representations. \review{The $\tau_c$ metric is used here because the Hidden Voices benchmark employs a 7-way rating system that is not directly aligned with the 11-way rating system used to train the model. $\tau_c$ evaluates the relative ordering of difficulty predictions, making it a suitable choice for comparing outputs across these misaligned systems.} Note that, as a benchmark dataset, the Hidden Voices assortment is only used for evaluation purposes, being the model exclusively trained on the PSyllabus one.

\begin{table}[h!]
\centering
\caption{Results in terms of the $\tau_c$ metric of the recognition performance of the model trained on the PSyllabus dataset and evaluated on the Hidden Voices benchmark.}
\resizebox{.9\columnwidth}{!}{%
\begin{tabular}{lccc}
\toprule[1pt]
\multirow{2}{*}{\textbf{Representation}} & \multirow{2}{*}{\textbf{Single}} & \multicolumn{2}{c}{\textbf{Multi-task}}\\
 \cmidrule(lr){3-4}
 & & \textbf{Era} & \textbf{Ranking}\\
\cmidrule(lr){1-4}
CQT & .59 (.02) & .63 (.04) & .63 (.03) \\
PR & .66 (.02) & \textbf{.67 (.05)} & \textbf{.67 (.03)}\\
\bottomrule[1pt]
\end{tabular}
}
\label{tab:zero}
\end{table}

As it may be observed, the findings in this experiment match those obtained in previous sections. More precisely, the PR representation stands as the most competitive one, especially when considering multi-task training schemes, which reports a correlation value of $\tau_c=0.67$ independently of the additional training task---era identification or additional difficulty ranking estimation---considered. Such an insight highlights the potential of multi-task training schemes for effectively decreasing the performance loss in zero-shot evaluation scenarios.

It must be highlighted that the presented insights in this work do not completely match those obtained when addressing the difficulty estimation task in sheet music images~\cite{ramoneda2023predicting}. While this is possibly due to the fact that the used dataset in this related work contained only 17 pieces, being, further research is needed to compare sheet music image classification audio to determine whether the modality and/or the quality of the datasets influences these generalization scores.

\section{A case study on multiple performances}
\label{case_multi}
This last experimental section of the work proposes a case study on leveraging different performances of a music piece---i.e., multiple recordings done by different performers----for improving the overall recognition results. The hypothesis in this case is that the decision-level fusion of the difficulty scores estimated for different performances of a given music piece may report better and more robust estimates than in the previous scenarios. The models presented in Table~\ref{tab:res_input_representations} are used exclusively for inference in these experiments. 

Formally, given a $P$-sized set $\mathcal{P}=\left\{x_1,\ldots,x_P\right\}$ containing different performances of the same music piece, we compute the collection of difficulty estimates $\mathcal{D}=\left\{\zeta(\hat{\mathbf{y}}_1),\ldots, \zeta(\hat{\mathbf{y}}_P)\right\}$ using the recognition model presented in \review{this} work. Out of these results we study different statistical indicators to integrate these individual estimations, being the ones evaluated the mean, median, and mode values. As a base indicator, we consider the random selection of \review{one of the $\mathcal{D}$ estimations}. \review{The absolute error between the expected label and the closest estimation refers to the smallest difference between the ground-truth difficulty level and any of the estimates in $\mathcal{D}$. This measure accounts for the best-case scenario where at least one prediction matches or closely approximates the ground truth.} The absolute error between the ground-truth label and the closest estimation is also provided for analysis purposes.

We considered the PSyllabus multiple performance benchmark dataset introduced in Section~\ref{benchmark_datasets}, selecting $P=5$ random interpretations per piece, being the results obtained for the previously discussed summarizing statistics provided in Table~\ref{tab:multi_performances_results}. The figures reported represent the average of 30 repetitions of the experiment with different seeds to ensure that the result does not depend on the random selection. Multi-task cases exclusively add the music era \review{as an} additional source of data.

\vspace{-0.5cm}
\begin{table}[ht!]
\centering
\caption{MSE, \review{Acc}, and $\tau_c$ results of the case studies on difficulty estimation with multiple performances.}
\label{tab:multi_performances_results}
\addtolength{\tabcolsep}{-.45em}
\resizebox{\columnwidth}{!}{
\begin{tabular}{llccccc}
\toprule
\multicolumn{2}{l}{\multirow{2}{*}{\textbf{Case}}} & \multicolumn{5}{c}{\textbf{Statistics}} \\
\cmidrule(lr){3-7}
 && \textbf{Random} & \textbf{Mean} & \textbf{Median} & \textbf{Mode} & \textbf{Closest} \\
\midrule
\multicolumn{7}{l}{\textbf{MSE} ($\downarrow$)} \\
\multicolumn{2}{l}{\textit{Single-task}}\\
& CQT & 1.5 (0.1) & \textbf{1.0 (0.1)} & 1.1 (0.1) & 1.3 (0.1) & 0.3 (0.1) \\
& PR & 1.3 (0.1) & 1.0 (0.1) & \textbf{0.9 (0.1)} & 1.0 (0.1) & 0.5 (0.1) \\
& MM & 1.3 (0.1) & \textbf{0.9 (0.1)} & 0.9 (0.1) & 1.1 (0.2) & 0.3 (0.1) \\
\multicolumn{2}{l}{\textit{Multi-task}}\\
& CQT & 1.6 (0.1) & \textbf{1.2 (0.1)} & 1.3 (0.1) & 1.5 (0.1) & 0.4 (0.1) \\
& PR & 1.3 (0.2) & 1.1 (0.1) & \textbf{1.1 (0.1)} & 1.2 (0.2) & 0.5 (0.1) \\
\midrule
\multicolumn{7}{l}{\review{\textbf{Acc}} ($\uparrow$)} \\
\multicolumn{2}{l}{\textit{Single-task}}\\
& CQT & 36.2 (1.2) & \textbf{38.2 (3.5)} & 37.4 (4.5) & 37.1 (3.6) & 75.0 (3.2) \\
& PR & 42.4 (2.4) & \textbf{45.2 (3.5)} & 44.9 (3.5) & 44.5 (3.1) & 71.6 (2.4) \\
& MM & 41.5 (2.0) & \textbf{45.3 (4.3)} & 44.3 (2.9) & 43.8 (2.8) & 74.4 (2.3) \\
\multicolumn{2}{l}{\textit{Multi-task}}\\
& CQT & 34.8 (2.5) & \textbf{38.8 (2.4)} & 38.2 (3.1) & 37.1 (3.7) & 71.5 (4.1) \\
& PR & 41.2 (3.4) & 44.2 (3.4) & \textbf{44.5 (4.0)} & 43.1 (4.2) & 70.1 (4.8) \\
\midrule
\multicolumn{7}{l}{$\mathbf{\tau_c}$ ($\uparrow$)} \\
\multicolumn{2}{l}{\textit{Single-task}}\\
& CQT & 0.8 (0.1) & \textbf{0.9 (0.1)} & 0.9 (0.1) & 0.8 (0.1) & 0.9(0.1) \\
& PR & 0.8 (0.1) & \textbf{0.9 (0.1)} & \textbf{0.9 (0.1)} & 0.9 (0.1) & 0.9 (0.1) \\
& MM & 0.8 (0.1) & \textbf{0.9 (0.1)} & 0.9 (0.1) & 0.9 (0.1) & 0.9 (0.1) \\
\multicolumn{2}{l}{$\text{Multi-task}$}\\
& CQT & 0.8 (0.1) & \textbf{0.9 (0.1)} & 0.8 (0.1) & 0.8 (0.1) & 0.9 (0.1) \\
& PR & 0.8 (0.1) & \textbf{0.9 (0.1)} & \textbf{0.9 (0.1)} & 0.8 (0.1) & 0.9 (0.1) \\
\bottomrule
\end{tabular}
}
\end{table}

The mean operator stands as the best integration policy, as it achieves the best recognition rates in most scenarios. The \review{median} operator, nonetheless, also reports its success in a number of cases, but with remarkable less proficiency than the former operator. The mode operator does not report an advantage in integrating individual estimates, as it does not outperform existing alternatives. 

Attending to the random case, it must be noted that all proposed policies improve this base case. Such a result shows that the particular integration policy does play a key role in the overall success of the task.

\review{Finally, it may be observed that the multi-task case does not provide a remarkable benefit to the task as the obtained figures are generally outperformed by single-task configurations.}

\vspace{-0.5cm}
\section{Conclusion}
\label{conclusions}

\review{In the context of music education, estimating the performance difficulty of a piece allows the creation of learning curricula appropriately adapted to the needs of each student.} Since manually estimating the difficulty of a music piece constitutes a tedious and error-prone task, researchers in the Music Information Retrieval field have presented different proposals to automate this process. Nevertheless, while successful to some extent, these schemes exclusively address high-level music abstractions such as machine-readable scores or music sheet images, hence neglecting the potential of directly analyzing acoustic recordings in this context.

This work addressed this gap in the field by introducing two main contributions: (i) the first audio-based difficulty estimation dataset---namely, \emph{Piano Syllabus} (PSyllabus) dataset---featuring 7,901 piano pieces across 11 difficulty levels from 1,233 composers as well as two additional benchmark datasets compiled for evaluation purposes; and (ii) a recognition framework capable of managing input representations---both in unimodal and multimodal manners---directly derived from audio to perform the difficulty estimation task. The comprehensive experimentation comprising a number of training strategies and evaluation scenarios not only shows feasibility of the task, but also proves the robustness of the model when addressing zero-shot recognition cases. All the code, models, and data are freely available for research purposes to encourage collaboration within the music education community.

\review{Future work plans to study the upper-bound performance of this difficulty estimation framework by providing the ground-truth transcriptions of the pieces. Representation learning techniques will be considered to better capture the piano performance characteristics prior to the difficulty estimation stage. Also, we aim at exploring the estimation this parameter at a finer level (e.g., music motif, section, etc) as an alternative to the current approach of a global score. Additionally, we plan to align the dataset with other modalities (e.g, symbolic music representations, sheet images and text descriptions) to create a multimodal assortment. We also aim to explore interpretability frameworks so that music educators may provide valuable insights for the curriculum development based on the model. Finally, we consider that involving educators in the task would contribute to the identification of its requirements and challenges, ensuring our solutions are practical and impactful.}

\section*{Acknowledgment}
The authors thank Ian Wheaton for his helpful guidance. We are also grateful to the Piano Syllabus web community for efforts devoted to compiling this syllabus as it constitutes the base resource of this research endeavour. We hope our work benefits this community as well as the music education one by aiding in the labeling of pieces and enabling the exploration of the forgotten cultural heritage.

We would also like to thank Nazif C. Tamer for suggesting the exploration of this task on acoustic recordings, as well as to Pablo Alonso and Oguz Araz for their insightful discussions.

This work was supported by ``IA y Música: Cátedra en Inteligencia Artificial y Música'' (TSI-100929-2023-1), funded by the Secretaría de Estado de Digitalización e Inteligencia Artificial, the European Union-Next Generation EU, under the program ``Cátedras ENIA 2022 para la creación de cátedras universidad-empresa en IA'' as well as the Generalitat Valenciana and University of Alicante through projects CIGE/2023/216 and INVA23-21, and Ministry of Education of the Republic of Korea and the National Research Foundation of Korea (NRF-2024S1A5C3A03046168).


\bibliographystyle{IEEEtran}
\bibliography{biblio}

\begin{IEEEbiography}[{\includegraphics[width=1in,height=1.25in,clip,keepaspectratio]{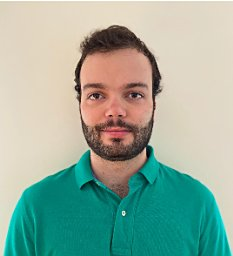}}]{Pedro Ramoneda}
Pedro Ramoneda holds a BSc in Computer Science from the University of Zaragoza, a Professional Degree in Piano Performance from the Conservatory of Music in Zaragoza, and an MSc in Sound and Music Computing from the Universitat Pompeu Fabra. He is currently a third-year PhD student in the Music Technology Group of the Universitat Pompeu Fabra under the supervision of Prof. Xavier Serra, focusing on the use of technologies from the Music Information Retrieval and Signal Processing field for supporting music education.
\end{IEEEbiography}

\begin{IEEEbiography}[{\includegraphics[width=1in,height=1.25in,clip,keepaspectratio]{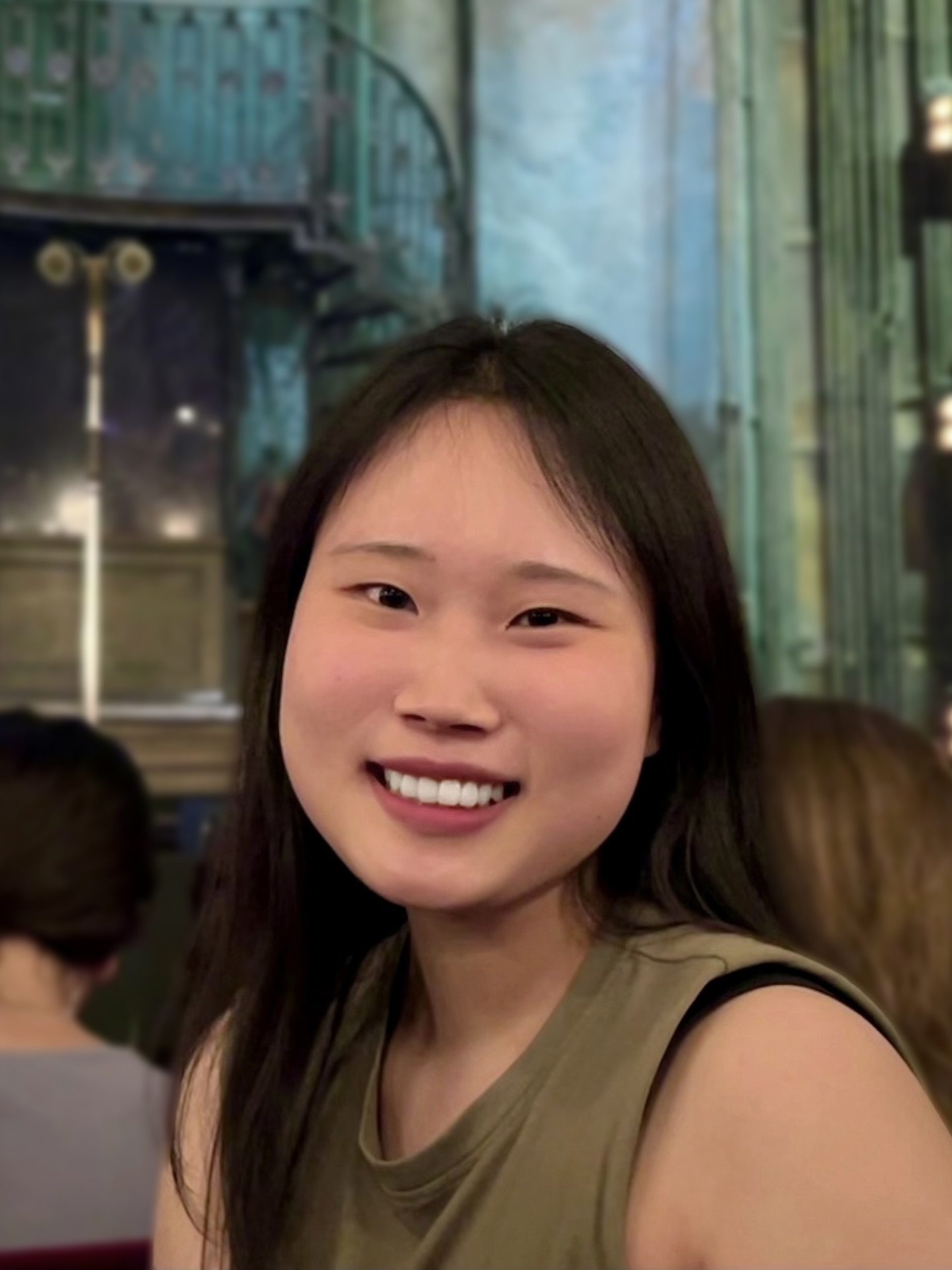}}]{Minhee Lee} is an undergraduate student pursuing a B.S. degree in the Department of Computer Science and Engineering at Sogang University in South Korea. She is currently an undergraduate intern in the MALer Lab under the supervision of Prof. Dasaem Jeong since 2023. Before joining the research group, she has done internships as a software engineer at Google in 2021 and 2022, and at FuriosaAI in 2022. Her research interest is on various music information retrieval tasks for understanding music with deep learning technologies. \end{IEEEbiography}

\begin{IEEEbiography}[{\includegraphics[width=1in,height=1.25in,clip,keepaspectratio]{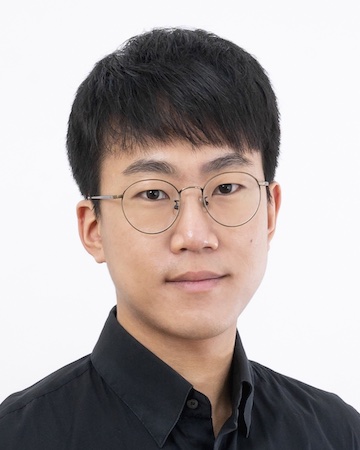}}]{Dasaem Jeong} is currently working as an Assistant Professor in the Department of Art \& Technology at Sogang University in South Korea since 2021. Before joining Sogang University, he worked as a research scientist in T-Brain X, SK Telecom from 2020 to 2021. He obtained his Ph.D. and M.S. degrees in culture technology, and B.S. in mechanical engineering from Korea Advanced Institute of Science and Technology (KAIST). His research primarily focuses on a diverse range of music information retrieval tasks, including music generation and musicology.
\end{IEEEbiography}

\begin{IEEEbiography}[{\includegraphics[width=1in,height=1.25in,clip,keepaspectratio]{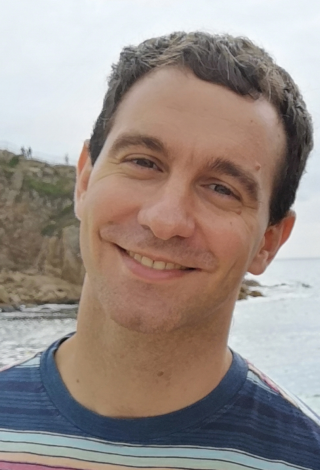}}]{Jose J. Valero-Mas} received his M.Sc. in Telecommunications Engineering in 2012 at University of Alicante, an M.Sc. in Sound and Music Computing in 2013 at Universitat Pompeu Fabra, and a Ph.D. in Computer Science in 2017 at University of Alicante. After three years as a data scientist, he was a Postdoctoral Researcher from 2020 to 2023. He is now an Assistant Professor at the University of Alicante, with research in Pattern Recognition, Machine Learning, Music Information Retrieval, and Signal Processing, and over 40 publications.
\end{IEEEbiography}

\begin{IEEEbiography}[{\includegraphics[width=1in,height=1.25in,clip,keepaspectratio]{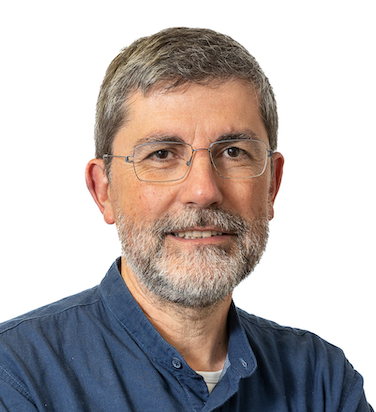}}]{Xavier Serra}
obtained his Ph.D. from Stanford and has pursued a career in music technology, focusing on the analysis, synthesis, and description of music signals. He integrates basic and applied research across scientific, technological, humanistic, and artistic disciplines. As the director of the Music Technology Group (MTG) for 30 years at Universitat Pompeu Fabra, he supervises research on sound and music signal processing and machine learning, with a focus on open science, to enhance the social and economic impact of his research.
\end{IEEEbiography}

\end{document}